\documentclass[structabstract]{aa}
\usepackage{txfonts}
\newcommand{\pq}{\ensuremath{P_Q}}
\newcommand{\pu}{\ensuremath{P_U}}

\usepackage{graphicx}
\begin{document}
\title{Polarimetry of transneptunian objects (136472) Makemake and (90482) Orcus
\thanks{Based on observations made at the La Silla-Paranal Observatory under programme ID 087.C-0615 (PI: S. Bagnulo)}}


   \author{I.N.~Belskaya \inst{1,2},
          S.~Bagnulo \inst{3},
          A.~Stinson \inst{3},
          G.P.~Tozzi \inst{4},
          K.~Muinonen \inst{5,6},
          Yu.G.~Shkuratov \inst{2},
          M.A.~Barucci \inst{1},
          S.~Fornasier \inst{1}}


\institute{LESIA-Observatoire de Paris, CNRS, UPMC Univ. Paris 06, Univ. Paris-Diderot, 5 Place J. Janssen, 92195 Meudon Pricipal Cedex, France\\
        \email{irina.belskaya@obspm.fr, sonia.fornasier@obspm.fr, antonella.barucci@obspm.fr}
\and Institute of Astronomy, Kharkiv National University, 35 Sumska str., 61022 Kharkiv, Ukraine
\and Armagh Observatory, College Hill, Armagh BT61 9DG, Northern Ireland, United Kingdom\\
\email{sba@arm.ac.uk, ast@arm.ac.uk}
\and INAF - Oss. Astrofisico di Arcetri, Largo E. Fermi 5, I-50125 Firenze, Italy\\
\email{tozzi@arcetri.astro.it}
\and Department of Physics, University of Helsinki, P.O. Box 64,  Gustaf H\"allstr\"omin katu 2a, FI-00014, Finland
\email{karri.muinonen@helsinki.fi}
\and Finnish Geodetic Institute, P.O. Box 15, Geodeetinrinne 2, FI-02431 Masala, Finland
}

   \date{Received 10 August, 2012 / Accepted 25 September 2012 }


  \abstract
   {We study the surface properties of transneptunian populations of Solar-system bodies.}
   {We investigate the surface characteristics of the dwarf planet (136472) Makemake and the resonant object (90482) Orcus.}
   {Using the FORS2 instrument of the ESO-VLT we have carried out linear polarisation measurements of Makemake and Orcus.}
   {Polarisation of Orcus is similar to that of smaller size objects. The polarimetric properties of Makemake are very close to those of Eris and Pluto. We have not found any significant differences in the polarisation properties of objects from different dynamical classes. However, there are significant differences in polarisation of large and smaller size objects, and between large TNOs with water-ice and methane-ice dominated surfaces.}
   {We confirm the different types of polarisation phase behavior for the largest and smaller size TNOs.  To explain subtle surface polarisation of Pluto, Makemake and Eris we assume that their surfaces are covered by a thin layer of hoarfrost masking the surface structure. }


\titlerunning {Polarimetry of Makemake and Orcus}
\authorrunning{I.N. Belskaya et al.}

   \maketitle
%

\section{Introduction}

The analysis of the polarimetric properties of transneptunian objects (TNOs) and Centaurs (Boehnhardt et al. 2004; Rousselot et al. 2005; Bagnulo et al. 2006, 2008; Belskaya et al. 2008a, 2010) has revealed so far many interesting results. In particular, two distinct behaviours of the polarisation phase curves have been found for transneptunian objects (Bagnulo et al. 2008). Objects smaller than 1000\,km exhibit negative linear polarisation that rapidly increases (in absolute terms) with increasing phase angle, and the observed polarisation reaches about $-1$\,\% at phase angles as small as 1$\degr$. By contrast, the largest TNOs exhibit small negative linear polarisation that does not change noticeably in the (limited) phase-angle range of the observations. These different types of polarimetric behaviour have been suggested to be related to different albedos and different capability of retaining volatiles for large and small TNOs (Bagnulo et al. 2008). The analysed sample of large TNOs have included four objects, namely Pluto (based on old photoelectric polarimetry by Breger and Cochran, 1982), and Eris, Haumea, and Quaoar (observed with the ESO VLT, Bagnulo et al. 2006, 2008, Belskaya et al. 2008a). The sample of small TNOs have included Ixion, Huya, Varuna, and 1999\,DE$_9$ (all observed with the ESO VLT, Boehnhardt et al. 2004, Rousselot et al. 2005, Bagnulo et al. 2006, 2008). 

In this paper, we continue to investigate the polarisation properties of the large TNOs and present the first polarimetric observations of (136472) Makemake and (90482) Orcus.

\section{Observations}

Polarimetric observations of two TNOs have been made in April-May 2011 at the ESO VLT. Linear polarisation has been measured with the FORS2 instrument (see Appenzeller et al. 1998) in the $R$ special filter. For Makemake, we have sampled nearly the full current observable phase-angle range (i.e., from $0.6\degr$ to $1.1\degr$), whereas Orcus has been observed only at a single phase angle of $1.1\degr$. Polarimetric observations and data reduction have been performed in the same way as in our previous observations (Bagnulo et al. 2006, 2008, Belskaya et al. 2008a, 2010). In order to derive the values of the reduced Stokes parameters \pq=$Q/I$ and \pu=$U/I$, we have used observations carried out with a half-wave retarder plate at all positions between $0\degr$ and $347.5\degr$  at $22.5\degr$  steps. The possible sources of instrumental errors were thoroughly investigated, and the detailed description of the reduction procedure can be found in Bagnulo et al. (2006).
\begin{table*}[ht]
\caption{Log of observations, measured polarisation parameters and R magnitudes}    
\label{table:1}      
       \scriptsize{
\begin{center}
\begin{tabular}{lllllrrl}        
\hline\hline                 
\multicolumn{1}{c}{Date}     &
\multicolumn{1}{c}{Sky}      &
\multicolumn{1}{c}{r}        &
\multicolumn{1}{c}{$\Delta$} &
\multicolumn{1}{c}{$\alpha$} &
\multicolumn{1}{c}{$\pq$}    &
\multicolumn{1}{c}{$\pu$}    &
\multicolumn{1}{c}{$R$}\\    
\multicolumn{1}{c}{(UT)}     &
\multicolumn{1}{c}{}         &
\multicolumn{1}{c}{(AU)}     &
\multicolumn{1}{c}{(AU)}     &
\multicolumn{1}{c}{($\degr$)}&
\multicolumn{1}{c}{(\%)}     &
\multicolumn{1}{c}{(\%)}     &
\multicolumn{1}{c}{(mag)}   \\
\hline                        
 Makemake\\
 2011 04 03.21&PHO&52.208&51.369&0.60&$-0.170$$\pm$0.045&$ 0.000$$\pm$0.045&16.81$\pm$0.05\\
 2011 04 25.06&PHO&52.202&51.517&0.80&$-0.316$$\pm$0.048&$-0.044$$\pm$0.048&16.86$\pm$0.05\\
 2011 05 10.06&CLR&52.213&51.676&0.94&$-0.050$$\pm$0.052&$ 0.089$$\pm$0.052&16.77$\pm$0.05\\
 2011 05 14.04&PHO&52.213&51.725&0.97&$-0.151$$\pm$0.085&$-0.093$$\pm$0.086&16.80$\pm$0.05\\
 2011 05 29.05&THN&52.216&51.926&1.07&$-0.114$$\pm$0.095&$-0.110$$\pm$0.087&\multicolumn{1}{c}{-}\\[2mm]
 Orcus\\
 2011 04 24.00&CLR&47.932&47.482&1.08&-1.062$\pm$0.092&-0.115$\pm$0.092&18.83$\pm$0.05\\
\hline
\end{tabular}
\end{center}
}
\end{table*}
The results of our observations are given in Table~1 that contains the epoch of the observations, the night time sky
conditions (THN \textendash  thin cirrus, CLR \textendash clear, PHO \textendash  photometric), the helio- and geocentric distances, the phase angle $\alpha$, the reduced Stokes parameters  \pq\  and \pu\ (expressed in a system with the reference direction perpendicular to the scattering plane), and the $R$ magnitudes. The Stokes parameters are thus expressed in such a way that \pq\  is equal to the flux perpendicular to the scattering plane minus the flux parallel to that plane, divided by the sum of the two fluxes. The parameter \pu\ characterizes the declination of the polarisation plane position from the plane normal to the scattering plane and, for symmetry reasons, is expected to be zero.
Acquisition images were used to measure the object magnitudes. In general, it is not possible to give accurate
night-by-night values for the zero point or extinction
coefficients because the photometric standard stars are not prescribed
under the FORS calibration plan for polarimetric
observations. Photometric standard stars have only been observed with FORS during two of the five nights in which our observation blocks have been executed. Therefore, we have used the  
nightly zero point and extinction coefficient available on the ESO  
Quality Control and Data Processing
web page. They have been calculated by using all the photometric
standard stars observed over a period of about 28 nights centered
at each night under consideration. We have associated an error of 0.05 mag
to the magnitude measurements which is consistent with the uncertainties
of the zero points. The errors due to photon noise and background subtractions are negligible in comparison with those of the zero points.

For each observing night, all polarimetric images have been stacked together to obtain a deep image as performed, e.g., by Bagnulo et al. (2010), and to search for possible activity. Our analysis did not reveal coma activity in any of these images.
\section{Results}
We plotted the measured polarisation of Makemake and Orcus and we compared it with the data available for other large (Fig. 1) and smaller-size (Fig. 2) TNOs. New observations confirm the different types of polarisation phase behavior of these two samples. Below we discuss implications of our observations for each object.

\subsection{(136472) Makemake}
Makemake is a dwarf planet with a diameter between 1360\,km and 1480\,km, assumed to have bright ($0.78 \le p_v \le 0.90$) and dark ($0.02 \le p_v \le 0.12$) terrains to fit thermal data (Lim et al. 2010). However, the lightcurve amplitude has been found to be very small ($\Delta V=0.029 \pm 0.002$). The most probable rotation period of the body is $7.771 \pm 0.003$\,h (Heinze and de Lahunta 2009). To make the small lightcurve amplitude of Makemake consistent with the assumption of dark spots, Lim et al. (2010) have proposed the following three possible explanations: (1) the dark spots are small and evenly distributed over the surface; or (2) the dark spots can be actually a band at constant latitude or a polar spot; or (3) the dark spot is an undiscovered satellite of Makemake. Further thermal measurements from Spitzer and Herschel space telescopes have confirmed unusual thermal emission spectrum of Makemake, which is different from all other large TNOs (M{\"u}ller et al. 2011). A recent stellar occultation by Makemake (Ortiz et al. 2012) has not excluded a slightly elliptical shape for this object. The average albedo has been estimated to be $p_v=0.71^{+0.08}_{-0.02}$ (Ortiz et al. 2012) which is in agreement with the estimation from the thermal data (Lim et al. 2010).
	Spectroscopic observations of Makemake have revealed strong absorption bands associated with methane (Barkume et al. 2005, Licandro et al. 2006, Brown et al. 2007, Tegler et al. 2007, 2008). Brown et al. (2007) have reported the presence of ethane features and no apparent evidence for N$_2$ and CO. Tegler et al. (2007, 2008) have reported a small shift of the CH$_4$ bands in their spectra of Makemake in comparison with a model fit, suggesting the presence of trace amounts of N$_2$ ice. The grain sizes of CH$_4$ ice in their modelling have varied from 0.1\,cm to 6\,cm and the best fit has been obtained for a model entailing grains of two sizes (Tegler et al. 2007, 2008).  Brown et al. (2007) have suggested that the main differences between the spectra of Pluto and Makemake are due to two main reasons: (1) depletion of nitrogen on Makemake's surface, which leads to large methane grains ($\sim 1$\,cm versus to $\sim$\,100\,$\mu$m for Pluto); and (2) presence of ethane ($\sim 100\,\mu$m grains) and tholin-like material. Eluszkiewicz et al. (2007) have proposed the slab surface model to avoid an assumption of unrealistically large methane grains. They have considered the absence of an opposition surge for Makemake as due to the presence of a slab surface.
	The main features of polarisation properties of Makemake can be summarized as follows:
\begin{enumerate}
	\item {} 
The linear polarisation is small with a mean value of $-0.16 \pm 0.05$\,\% at the phase-angle range of $0.6-1.07\degr$. In all five measurements, we have detected small negative polarisation, parallel to the scattering plane.
	\item {}  
We have not found any reliable changes of polarisation with phase angle and over the surface. Note that our measurements cover 70\,\% of the rotation cycle assuming the rotation period of 7.771\,h (Heinze and de Lahunta 2009). The values of $\pq=-0.32$\,\% measured at $\alpha$=0.8$\degr$ and $\pq=-0.05$\,\% measured at $\alpha = 0.94\degr$ probably reflect the scatter of our measurements rather than real changes in polarisation degree over the surface.

	\item {} 
The polarisation phase curve behaviour of Makemake is similar to that of Eris and Pluto, (Breger and Cochran, 1982). When we consider the values obtained at the same phase angles, the degree of linear polarisation measured for Makemake is closer to that for Eris than for Pluto and quite different from Haumea and Quaoar (see Fig.1).
\end{enumerate}

\begin{figure}[t]
      \centering
   \includegraphics[angle=0,width=7.7cm]{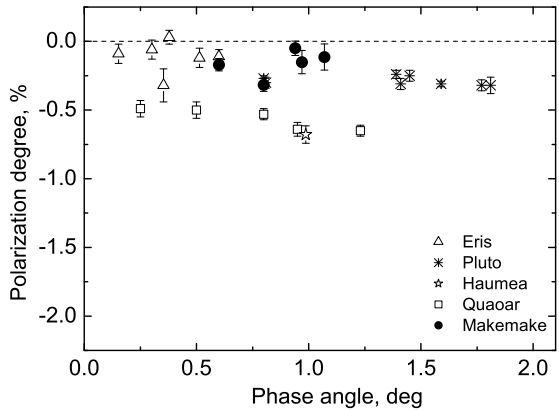}
\caption{Polarisation degree versus phase angle for Makemake (filled circles) and other large TNOs (Bagnulo et al. 2008, Breger and Cochran, 1982). }
\end{figure}

The polarisation properties of Makemake are similar to those of two other large TNOs with methane-dominated surfaces. The fact that the polarisation of Makemake is smaller (in absolute terms) compared to that of Pluto can be explained by the fact that Pluto is an unresolved system, and that the polarimetric measurements refer to the Pluto-Charon system. Assuming that Charon has intrinsic polarisation of about $-0.7$\,\% (similar to Quaoar and Haumea), we estimate that the intrinsic polarisation of Pluto is $-0.2$\,\% (see Eq. (1) of Bagnulo et al. 2008). The assumption that the polarisation properties of Charon are similar to those of Quaoar and Haumea is justified by the fact that all these bodies have water-ice dominated spectra (e.g. Barucci et al. 2011 and references therein). If this hypothesis is correct, the values of linear polarisation of the surfaces of Makemake and Pluto are very similar.
Our $R$-band photometry of Makemake in the phase range of 0.6-0.97$\degr$ is consistent with a shallow magnitude phase slope derived
by Heinze and de Lahunta (2009). They have found a linear phase coefficient of
$\alpha =0.037 \pm 0.013$\,mag/deg in the $V$ band for the phase-angle range of
0.59-0.84$\degr$ which is very similar to Pluto's phase coefficient of $0.032\pm0.001$\,mag/deg derived for the 0.4-1.8$\degr$ phase range (Buratti et al. 2003). In fact, the available photometry on Makemake, Eris, and Pluto can be fitted with Pluto's phase coefficient.
	Figure 1 shows that the polarisation properties of the largest TNOs having methane-ice spectra (Eris, Pluto, Makemake) are different from those of the TNOs with spectra dominated by water ice (Haumea, Quaoar). These differences are not related to the differences in their surface albedo. For instance, the geometric albedos of Makemake and Haumea are both ~0.8 (Stansberry et al. 2008), whereas the values of polarisation are $-0.16$\,\% and $-0.68$\,\% respectively.

\begin{figure}[h]
      \centering
   \includegraphics[angle=0,width=7.7cm]{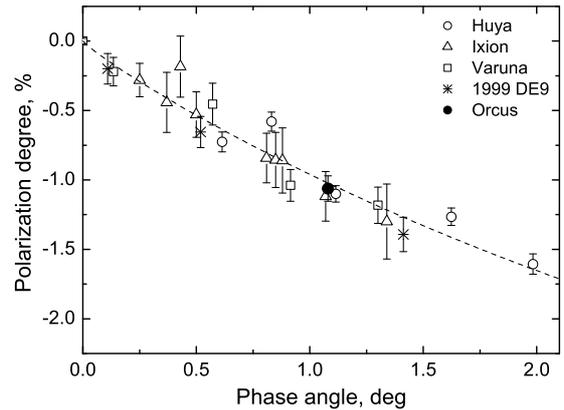}
\caption{Polarisation degree versus phase angle for Orcus (filled circle) and other TNOs smaller than 1000 km in diameter (data are from Bagnulo et al. 2006, 2008, and Boehnhardt et al. 2004).}
\end{figure}

\subsection{(90482) Orcus}
Orcus has a Pluto-like orbit in the 3$\colon$2 motion resonance with Neptune.
It is the largest known object among plutinos,
with diameter estimates varying between $850\pm70$\,km (Lim et al. 2010) and $940\pm 70$\,km (Brown et al. 2010). 
Orcus has a satellite, Vanth, about 2.6 mag fainter in $V$, and with a diameter of $280-380$\,km (Brown et al. 2010). Orcus has a water-ice-rich surface with an albedo of about 0.3 (Lim et al. 2010). Spectral modelling shows that water ice is presented mainly in its crystalline form (Carry et al. 2011 and references therein). Carry et al. (2011) have estimated an upper limit of about 2\,\% for methane and 5\,\% for ethane. Brown et al. (2010) have estimated a density of $1.5\pm 0.3$\,g\,cm$^{-3}$.
The linear polarisation of Orcus has been measured at the single phase angle of $1.08\degr$ (see Table 1). The measured value$\pq=-1.06 \pm 0.09$\,\% is in good agreement with the available data of other TNOs smaller than 1000 km in diameter (Fig. 2). Since Orcus is a binary object, we have estimated the intrinsic polarisation of Orcus using Eq. (1) of Bagnulo et al. (2008). Even if we assume that the polarisation intrinsic to Vanth is as large as $-1.5$\,\% at $\alpha=1\degr$, we find that the Vanth contribution to the observed polarisation is very small ($\approx 0.04$\,\%). Thus, the polarisation intrinsic to Orcus is quite different from that of Quaoar and Haumea (which have sizes larger than the size of Orcus), and it is instead rather similar to smaller size objects like two other plutinos Huya and Ixion.

\section{Discussion}
Our new observations have increased to 14 the number of TNOs and Centaurs for which polarimetric measurements have been carried out. The sample includes four dwarf planets, three Centaurs, three resonant, two scattered-disk, and two classical objects. We do not see any significant differences in polarisation properties of objects from different dynamical classes. Instead, there are significant differences in polarisation of large and smaller size objects, and between large TNOs with water-ice (Haumea, Quaoar) and methane-ice (Pluto, Eris, Makemake) dominated surfaces. Muinonen et al. (2012, in preparation) will discuss the diagnostic power of the numerical modelling of polarimetric measurements of these distant objects, which can be observed only within a very limited phase-angle range.  Here we discuss a few issues which should advance our understanding of the polarimetric properties of TNOs.

\subsection{Could the presence of a thin atmosphere be responsible for the small polarisation of Pluto, Eris, and Makemake? }
Breger and Cochran (1982) have hypothesized that the polarisation due to the atmosphere contributes to the observed polarisation of Pluto suppressing the polarisation from the surface. They have also noted that this hypothesis is consistent with a small discrepancy between the predicted and measured position angles of Pluto's polarisation found in 1972 (Kelsey and Fix 1973). However, successive polarimetric observations of Pluto in 1979-1981 have not revealed any systematic discrepancy in the position angles (Breger and Cochran 1982). Variations in the position angles exceeding the accuracy of measurements have been found neither for Makemake (see \pu\ in Table 1), nor for Eris (Belskaya et al. 2008a, Bagnulo et al. 2008). Pluto's atmosphere consists most probably of nitrogen with some carbon monoxide and methane with a pressure of a few microbars at the surface, when Pluto is near its perihelion (Stern and Trafton, 2008). Such a pressure is too small to warrant atmospheric effects on the surface polarisation. Even for Mars, with atmospheric pressures of several millibars, the influence is rather small in the visible spectral range (except for the contributions due to the clouds and the aerosol component). Since we can hardly expect the out-gassing for the dwarf planets to be larger than for Mars, we may exclude the atmospheric influence on the measured polarisation of Pluto, Eris, and Makemake. So the differences between large TNOs with methane-ice and water-ice dominated surfaces are related to the differences in the physical properties of their topmost surface layers.

\subsection{What is the most probable cause of the negative polarisation of TNOs?}
The negative polarisation of a regolith-like surface
can be formed by several physical mechanisms.
Two of them, namely, the single scattering by particles and the coherent multiple scattering,
are considered to be the most important mechanisms for planetary regolith (e.g., Shkuratov et al. 1994, 2002, Muinonen 2004, Muinonen et al. 2011).
The single-particle scattering produces broad negative polarisation branches and have a minor contribution at small phase angles. The mechanism of coherent backscattering is effective for bright surfaces. It produces a narrow branch of
negative polarisation usually accompanied with a narrow brightness opposition effect. Such narrow surges both in polarisation and brightness, typically centred at $\alpha \sim 0.4-1\degr$, have been found for high-albedo planetary satellites and asteroids (see Mishchenko et al. 2010 for a review). The relationship of polarisation degree and albedo may help to distinguish between different mechanisms of the negative polarisation. Figure 3 shows the albedo dependence of polarisation degree at a phase angle of 1$\degr$ for TNOs and Centaurs. There is an inverse correlation similar to the correlation of $P_{\rm min}$ and albedo measured for regolith-like surfaces and expressed as
log~$p_v$ = $C_1$log($P_{\rm min}$) + $C_2$,
where $C_1$ and $C_2$ are constants. The linear fit in a log-log scale to the data for TNOs and Centaurs gives $C_1$=1.00$\pm$0.03 and $C_2=-0.92\pm 0.02$ which are close to the values adopted for asteroids $C_1=1.22$, $C_2=-0.92$ (Lupishko and Mohamed 1996). The similarity of these dependences assumes that the polarisation of TNOs measured at a phase angle of 1$\degr$ is probably close to $P_{\rm min}$; i.e. the minimum of polarisation of TNOs is shifted to small phase angles as compared to asteroids. New observations confirm our previous conclusion that a single measurement of polarisation degree of a TNO at phase angle of 1$\degr$ can provide at least a distinction between high and low albedo surfaces (see Bagnulo et al. 2008).
\begin{figure}[h]
      \centering
   \includegraphics[angle=0,width=8cm]{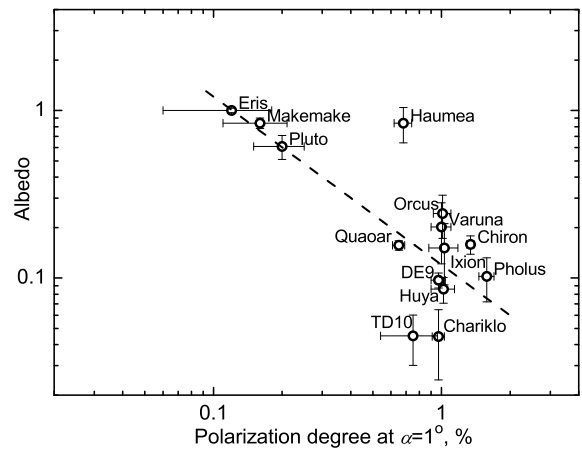}
\caption{Dependence of polarisation degree measured at phase angle of 1$\degr$ on albedo in log-log scale. The line corresponds to the asteroid relation of $P_{\rm min}$ vs. albedo. Albedos have been taken from Stansberry et al. (2008), Sicardy et al. (2011), and Lim et al. (2010).}
\end{figure}
The inverse correlation of negative polarisation and albedo can be explained by an increase of incoherent multiple scattering for bright objects which suppresses polarisation due to single particle scattering. If the main mechanism of negative polarisation of TNOs is the coherent backscattering we can expect deeper polarisation for brighter objects. However, Eris, Makemake and Pluto show the smallest values of polarisation. An exceptional case is Haumea, for which the measured polarisation degree deviates from the general trend (see Fig.3). We could assume higher contribution of the coherent backscattering for Haumea, but its magnitude phase curve is shallow down to 0.5$\degr$ and does not show an opposition surge (see Rabinowitz et al. 2006). New polarimetric observations of Haumea are needed to understand possible reasons of its deeper negative polarisation as compared to other bright TNOs.

\subsection{Why we do not measure the opposition surges for the brightest TNOs predicted by the mechanism of coherent backscattering? }
Polarimetric measurements of Eris made in the phase angle range of
0.15-0.5$\degr$
has not revealed any opposition surge (Belskaya et al. 2008a).
The albedo of Eris is found to be very high: $p_V$=0.96 (Sicardy et al. 2011).
Our observations of Makemake have not shown any opposition surge at
$\alpha$= 0.6-1.1$\degr$ for this high-albedo object ($p_V$=0.7-0.8).
Most probably, the opposition surges are extremely narrow and not covered by
the available observations. Unfortunately, it is impossible to verify this
 suggestion in the nearest future due to the inaccessibility of very small phase
 angles for these objects. We note that recently, a very narrow opposition surge
 at $\alpha$$\textless$0.1$\degr$ was found for the neptunian satellite Triton
(Buratti et al. 2011). Triton is one more object with a methane-rich surface
 (e.g., Tegler et al. 2012) and is considered to be a captured Kuiper-belt object. This also supports our suggestion of the existence of narrow backscattering peaks for Makemake, Eris, and Pluto.

\subsection{Which surface properties can be responsible for very narrow backscattering surges? }
Laboratory measurements at very small phase angles (Psarev et al. 2007) have shown that narrow opposition spikes are inherent for bright and fluffy surfaces consisting of very small particles. The narrow spikes are usually accompanied with narrow branches of negative polarisation. This is confirmed by laboratory polarimetric measurements of the fluffy SiO$_2$ sample (Shkuratov et al., 1994, 2002; see also the numerical results by Muinonen and Videen, 2012). The width of the polarimetric spike depends on the porosity of particulate surface. An example of such measurements is shown in Fig.~4 for the same SiO$_2$ sample before and after compressing that significantly change the surface porosity.
\begin{figure*}
      \centering
   \includegraphics[width=17cm]{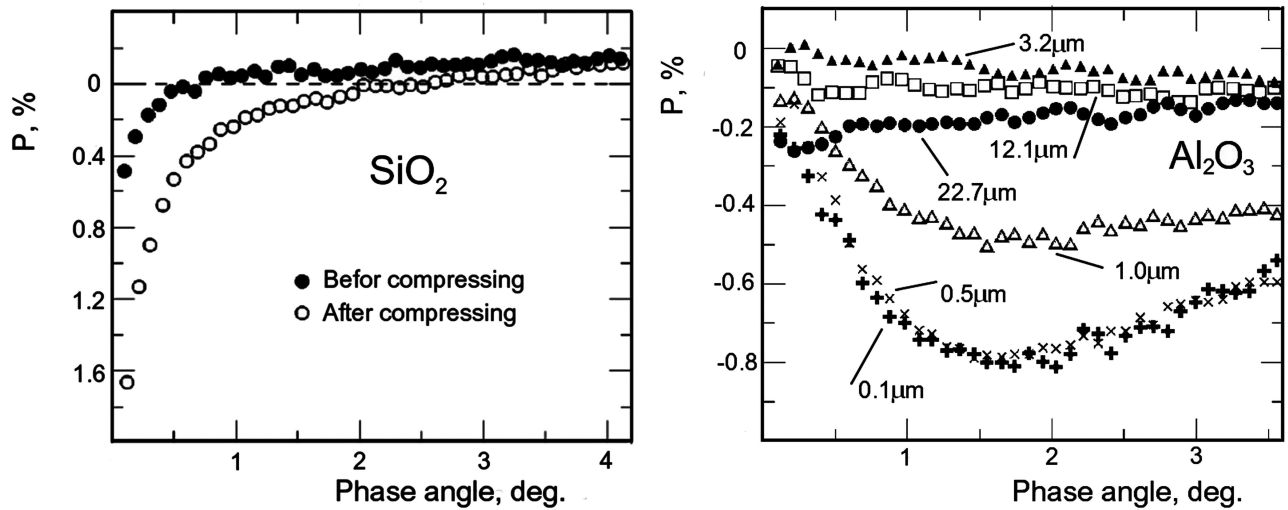}
\caption{Polarisation phase curves for a superfine powder of SiO$_2$ before and after compressing with alcohol drying at
$\lambda =0.63\,\mu$m (left panel) and for size-separated Al$_2$O$_3$ powders at $\lambda =0.63\,\mu$m (right panel). These figures are taken from Shkuratov et al. (2002).
}
\end{figure*}
Another possible explanation of the absence of the conspicuous polarimetric spikes is that surface particles are large enough compared to the wavelengths of visible light (Belskaya et al. 2008a). Figure 4 illustrates this with laboratory polarimetric measurements of $Al{_2}O{_3}$ powders with different average sizes of particles and almost the same high albedo (Shkuratov et al. 2002). One can see that powders consisting of large particles show a shallow branch of negative polarisation.

What is the most realistic model for the surfaces of Makemake, Eris, and Pluto? Are the surfaces formed by large ($\sim 10\,\mu$m) particles, or by tiny ($\le 0.1\,\mu$m) particles that create a very porous (fairy castle) structure? The latter seems to be more realistic, as the surfaces of Makemake, Eris, and Pluto are composed of frozen gases (nitrogen, carbon monoxide and methane) that may form porous hoarfrost from their diluted atmospheres. One may expect that the atmospheres can freeze up seasonally as the bodies move further away from the Sun. The $N{_2}$ and $CH{_4}$ frosts evaporate preferentially on the darker (warmer) regions and refreeze on the brighter (colder) regions,
producing atmospheric transport of volatiles. This can also be related to possible sub-surface activity.

\section{Conclusions}

Polarimetric observations of Makemake and Orcus confirm the different types of polarisation phase behaviour of the large and smaller size TNOs. Polarisation of Orcus is similar to that for smaller size objects. Polarimetric properties of Makemake are very close to those of Eris and Pluto, and different from those of Haumea and Quaoar. To explain the small polarisation of Pluto, Makemake, and Eris we assume that their surfaces are covered by a thin fluffy layer of hoarfrost masking the "ordinary" surface structure that might produce prominent negative polarisation branches.

\begin{acknowledgements}
This work has been supported by the COST Action MP1104
entitled "Polarisation as a tool to study the Solar System and beyond",
by the NASA program for Outer Planets Research (contract NNX10AP93G)
and by the Academy of Finland (contract 127461).
\end{acknowledgements}

\end{document}